\documentclass[a4,12pt]{article}
\usepackage{epsf}                                
\usepackage{graphicx}
\usepackage{amsbsy,amssymb,latexsym}

\catcode`\@=11
\@addtoreset{equation}{section}

\global\arraycolsep=2pt
\begin{document}

\begin{flushright}
{
YITP-06-31\\
OIQP-05-16\\
hep-th/0612156
}
\end{flushright}
\bigskip

\begin{center}
{\Large 
Entropy Currents for Reversible Processes in \\
a System of Differential Equations.\\
-- The Case of Latticized Classical Field Theory --\\
}
\vfill

Holger B. Nielsen \\ 
{\it
Niels Bohr Institute, \\ 
University of Copenhagen\\ 
17 Blegdamsvej, Copenhagen, $\phi$, Denmark
}
\\
and
\\
Masao Ninomiya 
{\it
\footnote{Working also at Okayama Institute for Quantum
Physics, Kyoyama-cho 1-9, Okayama City 700-0015, Japan.}\\ 
Yukawa Institute for Theoretical Physics, \\
Kyoto University \\ 
Kyoto 606-8502, Japan}
\end{center}

\bigskip

\begin{abstract}
We consider a very complicated system of some latticized 
differential equations that is considered as equations of motion for a field
theory.  
We define macro state restrictions for such a system
analogous to thermodynamical states of a system in statistical
mechanics.  
For the case in which we  have assumed adiabaticity in a generalized way 
which is equivalent to reversible processes.
It is shown that we can define various
entropy currents, not only one.  
It is indeed surprising that, for a two dimensional example 
of lattice field theory, 
we get three different entropy currents, all
conserved under the adiabaticity condition.  
\end{abstract}

\section{Introduction}

In classical mechanics one can define entropy by specifying 
macro states, $A$, as Boltzmann's constant times the logarithm of the 
phase space volume of the space of micro states corresponding to such a 
macro state:
\begin{equation}
S(A) = k \log(\mbox{PS vol}(A)) .
\end{equation}
Here $A\subseteq$ Phase Space, (PS for short), is the set of all the micro states
$(\vec{q}, \vec{p})$ that agree with the macro state $A$;
\begin{equation}
A = \left\{(\vec{q}, \vec{p}) \mid (\vec{q}, \vec{p}) \in A\right\} .
\end{equation}

We suppose that we can distinguish various ``macro states'' 
in the sense that
collections (subset, roughly speaking) of states of the system at the
fundamental level correspond to what we collect under one state in
macroscopic thermodynamics.  
If we can assign these macro states
entropies, we can define them as logarithms of the number of
states in the macro state, according to usual definition 
\cite{1}\cite{2}\cite{3}.  
This ``number of states'' in the macro
state may be able to be defined in quantum mechanics, but in a classical
description we either have to replace the number of states by a volume
of the corresponding phase space or we have to discretize the
phase space.

When the system is described macroscopically and behaves adiabatically, 
we shall seek to define a kind of currents very generally as below.
The point $(\vec{q}, \vec{p})$ is phase-space point = micro state.

Usually one specifies a reversible development of the macro state by taking
it to be very slow.
This means that one can invert the rate of development of the macro state in time.
If there is no such fact, it may cause significant influence on the micro
physics and micro development to the second order in the development rate of the 
macro state $\dot{A}^2$.
Combining this approximation with the second order effect of $\dot{S}\left(A(t)\right)$ in the second law of thermodynamics $\dot{S}\left(A(t)\right) \ge 0$,
requires
\begin{equation}
\frac{\partial S}{\partial A} = 0
\end{equation}
which is nothing but conservation of $S$, if we use a simple formal 
Taylor expansion
\begin{equation}
\frac{d}{dt}S\left(A(t)\right) = \frac{\partial S}{\partial A}\dot{A}(t) 
+ \frac{1}{2} \frac{\partial^2 S}{\partial A^2}\dot{A}^2(t) + \cdots .
\end{equation}
Here we neglect all but first term in the slow limit.

Part of the motivation of the present article is the hope to make more
general definition of reversibility criteria.

A direction in which such generalization has already been done in many
articles and textbooks.
We also go and seek to construct entropy distribution in space and entropy
current, so to speak answer where the entropy is.

We would contemplate to specify in which field variables would sit the 
entropy in a latticized field theory.
In such a latticized field theory one has collections of variables and
equations of motion.
One could hope to obtain a general description as to how the entropy flows
under certain restrictions of macro state.

In the long run one might seek to study entropy flow in a Euclidean field theory
(latticized or not).
In studying this it is most elegant and easy to work in the case of reversible
processes because the entropy gets conserved.
In such processes $\dot{S} =0$ holds so that the second law of thermodynamics is
fulfilled in a trivial manner and furthermore time reversal invariance too.

Therefore this second law is not very relevant if we restrict ourself to 
reversible processes as is actually the case in the present article.
This touches on a motivation of ours for present work:
We study entropy flow without necessarily assuming second law at the outset.

This paper is the first attempt in a series of papers on this entropy flow subject.
We then expect that the entropy will be conserved.  
In fact in cases where we have
latticized field theories we would expect that we could
find a conserved current $j^\mu_s(x)$ of entropy
\begin{equation}
\partial_{\mu}j^{\mu}_s(x)=0.
\end{equation}

It is the main purpose in the present article to discuss and construct
such an entropy current under an abstract and general set up.
However we investigate in detail 
only in a couple of examples the most important of which is a
certain triangular lattice in two space time dimensions.

We shall meet in this article a few surprising results in connection
with defining such entropy currents for the latticized field theory models:  
It turns out that in two dimensional space time
we actually have to define \underline{three}
different entropy currents $j^\mu_A(x)$, $j^\mu_B(x)$ and
$j^\mu_C(x)$ rather than just one as one might have a priori expected.

When one talks about entropy 
it is normal to think about the second law of thermodynamics tells that entropy
\cite{4} will always increase or stay constant, but never fall down.
If we, however, as in this article, restrict ourselves to the adiabatic process
the entropy gets constant and the second law is not so relevant anymore.  
In fact it is realized trivially.
We would rather like in the present work to think of it under the condition 
that we have imposed an abstractly defined adiabaticity \cite{5} condition .  
Once this is assumed to be
valid for our model we will be able to find the conserved entropy
current without having to make any use of the second law of thermodynamics.  
We therefore would like to think of
our calculations as performed in a world without
having any second law of thermodynamics at all.  
In other words we should rather
think of the present article as a work relevant for working
prior to the second law of thermodynamics.  
Thus we should be able to
make use of our considerations in an attempt to derive the second law
from physics assumed at a more fundamental level.  
We must though admit that in the present article we assume adiabaticity:
It is defined
though in a so abstract manner that we do not need second law for that either.  
Thus our considerations could be especially relevant for models
with compact space time for which non-trivial relevance of second law
of thermodynamics may not be possible.

Since we only consider trivial realization of the second law of thermodynamics,
$\dot{S}=0$, the time orientation associated with this second law becomes quite
relevant. 
Thus
we also do not really use the metric tensor for space time and 
through most of the present article it is not specified whether we use 
Minkowskian or Euclidean metric.  
However the lattice which we use in the example is a two dimensional triangular 
one and it has a $120^\circ$ degree rotation symmetry.  
This symmetry suggests a
Euclidean metric since it is a subgroup of the Euclidean rotation.

The lack of specification of a time axis orientation immediately
calls attention to the fact that it needs a convention of sign 
in order to define an entropy density $j^o_s(x)$ which is part of 
an entropy current $j^\mu_s(x)$.  

The present paper is organized as follows:  
All through the article we shall make use of the concept of macro states or 
rather macro restriction in the sense of a subset of all the micro states.
There are fundamental field configurations, corresponding to a class
of such states conceived as not distinguished in the macroscopic description.

In section 2 we introduce our two dimensional classical lattice field theory.
In section 3 we describe the restriction on the field $\varphi$.
In section 4 argue a reversability (or adiabaticity) requirement as a local
principle of no loss of micro solutions by local interplay.
In section 5 we give a one dimensional example that illustrates our formalism.
In section 6 we then go to our main example, the two dimensional latticized theory.
In section 7 we then define entropy currents $j^\mu_A$, $j^\mu_B$ and $j^\mu_C$
associated with each ``class of half curves on the lattice".
In section 8 we connect these quantities with our entropy flow currents by 
some equations.
In section 9 we emphasize that we obtained three entropy currents.
In section 10 we point out two alternative way of constructing our three entropy
currents.
In section 11 we then actually construct one of them, $j^\mu_A$ using 
conservation law in our formalism.
In section 12 the continuum limit is considered for this current $j^\mu_A$.
In section 13 the formalism is described analogously for the rest two currents
$j^\mu_B$ and $j^\mu_C$.
In section 14 a relation between the three entropy currents is presented.
In section 15 we seek to understand three entropy currents by relating them to
numbers of solution.
Section 16 is devoted to present conclusion and outlook.

\section{Introduction of lattice}

We want to study reversibility hypothesis
tells us for a field theory with space so that
we can study how entropy may flow.

For simplicity we shall exercise by a two dimensional space-time
world and even discretize it into a lattice.

We may at first set up the lattice theory quite naively by assuming
that we have it as a system of first order differential
equations in which the derivatives are discretized and made into differences.  
We may think of the discretized derivatives to be one-sided; 
for instance we may choose the forward difference equation to the discretized differential
one
\begin{equation}
\frac{\partial }{\partial x}f(x) \longrightarrow  \frac{1}{a}\left\{f(x+a)-f(x)\right\},
\end{equation}
where $a$ is a lattice constant that can be set to one and $f(x)$ a function.
We shall end by having the differential equations turned
into an equation set, one per site.  This equation will involve
the fields on three different sites, namely
\begin{enumerate}
  \item[1)] on the selected site 
  \item[2)] on the site following the above first one in the x-direction
  \item[3)] on the site following the second one in the y-direction.  
\end{enumerate}

It is the latter two sites that are needed to make the discretized
derivatives.
In d-dimensional space there will be analogously a need for involving
at least $d+1$ lattice sites in the equations.
It is important for having the number of equations correctly that we
have just one such $d+1$ (=3 in $d=2$ case) equations for each site.
We may draw symbolically as Fig. 1 the involvement of the sites in the equations
by encircling the sites involved in an equation.  Then the just
described set up comes to look as Fig. 1.

\begin{center}
\includegraphics[width=6cm]{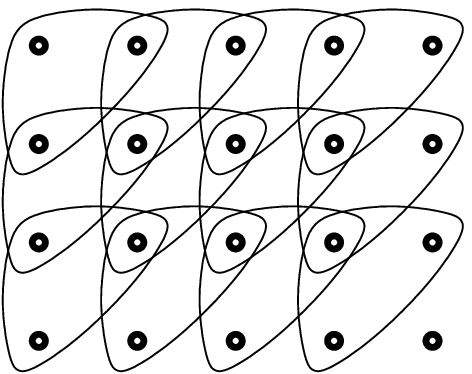}
\begin{figure}[h]
  \caption{Field equations on a lattice}
\label{fig:e1.eps}
\end{figure}
\end{center}  
\newcommand{\DELTA}{{\raisebox{-0.1em}{\includegraphics[width=0.4cm]{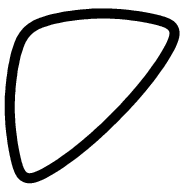}}}}
There is indeed the same number of sites 
\includegraphics[keepaspectratio=true,height=3mm]{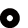}'s
representing fields
per unit area 
as the number of equations of motion 
as depicted in Fig. 1.
The encircling lines \DELTA
represent the field equations.  
So if we had a compact
two dimensional space - a torus say - covered by this pattern there
would be equally many variables as equations.  
The \DELTA- symbols represent equally many equations as the field components.  
Thus we expect generically that we get an equation system with
just a discrete set of solutions.  At first you might expect this
equation system to have just of order of unity solutions.  
As we already know there is much more than of order of 
unity solutions for systems behaving chaotical behavior.

We might draw more elegantly the lattice by deforming it so as to draw
as Fig. 2 in a hexagonally symmetric way:
\begin{center}
\includegraphics[width=10cm]{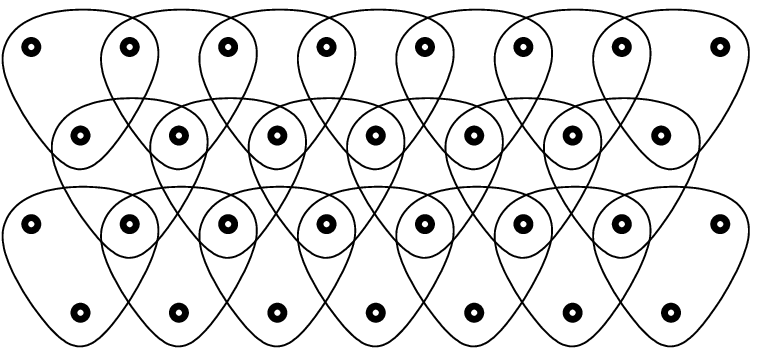}
\begin{figure}[h]
  \caption
  {{\small Field equations on a lattice with a hexagonal symmetry deformed from Fig.1}}
\end{figure}
\end{center}

\section{Formulation of equation of motion as a relation between field values on lattice.}
Suppose we have a lattice of points with some fields defined on
each of these points.
We may be so abstract that there may not even
be the same number of components and type of fields at each lattice
point, but such an extension will not play important role in this section.
We can give a name to the set of field configurations F.  
Then a field development through all times
of the whole system is described at the micro level by a function
$\varphi$ defined on the lattice L  and taking values in F.  
That is to say
\begin{equation}
\varphi:L\to F.
\end{equation}
If we want to be very general and let $\varphi$ take values that are
specific for the various sites $s\in L$ we can just formally put the
value set $F_s$ allowed for $\varphi(s)$ into $F$ as subset and write
$\varphi:L \to F$ with restrictions
$\varphi(s)\in F_s \subseteq F$.

We then introduce the concept of a ``macro restriction''.  
The idea is that it generalizes the restriction that
some part of the world at some time in some era is in a special macro
state in the thermodynamical sense.

The macro restriction is meant to be the restriction on the micro degrees of freedom,
i.e. the field $\varphi$ at the lattice points which ensures this micro configuration 
to be conceived as the macro state.
It is taken to define the ``macro restriction" in question.
By generalizing this concept of macro (state) restriction, we would like to
introduce a more broad concept of restriction between field values on the lattice points.
Especially it should be possible to have such restrictions between field configurations in
successive time moments and thus we could consider the equations of motion as a 
special case of a restriction.

The macro restriction is now defined to be a constraint on the
function $\varphi$ on a subset, $B$ say, of the lattice $L$.  That is
to say the ``macro restriction'' is defined by specifying a subset of
functions
\begin{equation}
R \subseteq\{\chi:B \to F  {\rm ~with} ~\chi(s) \in F_s \}.
\end{equation}
We shall be mainly interested in ``local macro restrictions'' by which we
mean that the lattice points of $B$ lies in the neighborhood of each
other.

Then we think about a possible macro description as a set of macro
restrictions $\{R_i \mid i \in I\}$ 
where I denotes the set of all positive integers.
Thus we require about an ``allowed''
map
\begin{equation}
\varphi :L \to F \hspace{5mm} (\varphi(s) \in F_s)
\end{equation}
that it should obey
\begin{equation}
\forall i\in I [\varphi \mid_{{\rm restricted ~to} ~B_i} \in R_i].
\label{3.4}
\end{equation}
The easiest to think about is really to think of the $F_s$'s (or $F$)
as  discrete countable or even finite sets, so that we simply can count
the number of functions of the type
\begin{equation}
\varphi:L\to F \hspace{5mm} (\varphi(s)\in F_s)
\end{equation}
corresponding to a given ``macro description'' $\{R_i \mid i \in I \}$.

We have to keep in mind that such a macro description is a description from a
macroscopic point of view of the time development of some 
classical field theory.  The lattice $L$ is a space time lattice
so that each lattice point represents both a space point and a moment
of time.

The latticized equation of motion would mean - typically for a first
order differential equation - a relation between $d+1$ neighboring
lattice point fields.  Here $d$ is the number of space time dimensions.
Such restrictions $\varphi$ of equation of motion to a subset of $d+1$ neighboring
points $B_i$ are formally of a completely similar form as the macro restrictions
(\ref{3.4})
The function subsets $R_i$ connected with the equations of motion are of a 
special type.
In fact in the simplest case
\begin{equation}
CnB_i = d+1
\end{equation}
for these equations of motion restriction.
Here cardinal number $Cn$ is given by $CnB =$ the number of elements in $B$.
That is to say we can consider the equation of motion as a lot of
macro restrictions with the $B_i^{,s}S$ involving $d+1$ lattice point 
i.e. $ktB_i=d+1$, where $kt$ means cardial number of the set.  
The set $R_i$ for the equations of motion which represent macro
restrictions are then subsets of dimension $d\cdot n$ of the full
$(d+1)\cdot n$ dimensional space on the space of all functions defined on the
subset $B_i$ of the lattice.  Here we denoted by $n$ the dimension of
$F_s$'s - assumed to be the same for all $s\in B_i$ - so that the $n$
equations of motion just bring the dimension of the space of functions
$\varphi $ restriction to $B_i : B_i \to F_s$ down from $(d+1)\cdot n$ to
$d \cdot n$.

\section{Principle of no loss of micro solutions\\
developments by local interplay of restrictions}
In this section we want to formulate the requirement of  only reversible
(or adiabatic) processes.

We have recently pointed out \cite{6} that if one imposes a specific periodicity
on a mechanical system, then the system is generically forced to behave reversibly.
Thus we could use such a special model with imposed periodicity that the model 
has compact space time as a simple example of a model in which reversibility is
imposed.
But let us stress that this is just an example and that we in general we assume that 
we work with reversible processes only.

As the simplest case we want to consider a compactified space time so that there is 
no such thing as $t$ or the space coordinate running off to infinity.

Since in this case we expect very few solutions if any, it may sound a little strange
to make statistical considerations concerning these solutions.
Let us, however, at first think in this statistical way and imagine that for a given
system of macro state restrictions one can ask for a probability for solutions there.

For many combinations of macro restrictions we really risk that that probability for 
at least one solution would be extremely low.
One might assign even meaning to low probability for getting a solution because 
it would occur with a tiny probability measure in the parameter space.

In our article we actually introduced a way of working with generic equations of motion
by taking parameters in the Hamiltonian as random numbers, so that one could 
in principle ask for probabilities of getting various numbers of solutions.

Naturally we should consider the case that the higher the number of allowed
functions $\varphi$ (``solutions") the more likely is a certain system of macro
restrictions, what we called a macro description.  
So we should be most interested in macro scenarios with the largest number of allowed
functions $\varphi:L \to F$.  We can say in this way when we have
taken $F$ or $F_s$'s discretized form so as to be able to simply count functions
$\varphi$.  
However, we may treat quite analogously the cases in which we
think of $F$ as a space with 
continuous coordinates on it, provided it is possible at the end to define a measure
on the set of solutions $\varphi$ that obey the equation of restrictions.
Then such measure may replace the counting in the totally discrete case.

Typically $F$ would be phase space with
a Liouville measure on it, but for pedagogical reasons we shall 
at first consider the discretized case in which we can simply
count the number of functions $\varphi$.  
If we had not made the
space on which the lattice $L$ is distributed a compact one the lattice
$L$ might be infinite and we may for that reason also get problems with
pure counting.  
Since we shall, however, be most interested in small
local regions we shall be satisfied with counting; we first of
all look for pieces of the lattice $L$ with only a finite number of
sites.

We should impose the following principle on the system of restrictions:
Whenever the sets of lattice points $B$ and $D$ involved in two restrictions
$(R_B, B)$ and $(R_D, D)$ have a non-trivial overlap $B\cap D$, then the
restrictions here will be consistent in the sense that the imposition of one
of the restrictions must not reduce the number of field configurations allowed
on the overlap $B\cap D$.

We may have to keep in mind that if we have in certain macro scenario two
macro restrictions defined on the subsets $A$ and $B$ of the lattice
with $A\subseteq B$ then we can replace those macro restrictions by a
single one ``$R_A \cap R_B$''on the bigger subsets of the ones of $L$,
namely $B$.  That is to say we can choose a new restriction defined on $B$:
\begin{eqnarray}
``R_A \cap R_{B}"=\{ \varphi \mid_{{\rm restriction ~to}~B} \mid 
& \varphi \mid_{{\rm restriction ~to}~B} \in R_B \wedge\nonumber\\
& \varphi \mid_{{\rm restriction ~to} ~A} \in R_A \}
\end{eqnarray}
We can thus reduce our considerations to macro scenarios built up from
a set $\{(R_i,B_i)\mid i \in I\}$ where non of the $B_i$'s are
contained in any other one.

If two such restrictions, say $(R_D,D)$ and $(R_B,B)$ have a
non-empty intersection of their lattice sets
$D\cap B\neq\emptyset$, then we can ask for whether the restriction
of $\varphi$ to the overlap region $\varphi\mid_{\rm restricted ~to ~B \cap D}$ is allowed to be the same set of restricted functions for
$(R_D,D)$ as for $(R_B,B)$.  
If the $\varphi\mid_{\rm restriction ~to ~B \cap D}:B\cap D\to F$ 
functions allowed by $(R_D,D)$ and
$(R_B,B)$ are not essentially the same then the functions over $B$
belonging to the set $R_B$ which are also allowed by $(R_D,D)$ will be
reduced in number relative to the number of functions in $R_B$.  

Under the statistical way of which we would like to justify by some
physical considerations above in section 3, 
there is only a
fraction of the set of functions $R_B$ that can be realized. 
That will at the end reduce the number of allowed $\varphi$ by the ratio
giving the fraction of $R_B$ which get allowed by $(R_D,D)$.  By having
such a lack of match on the overlap region $B\cap D$ we have basically
lost a fraction of the potentially achievable allowed functions
$\varphi$.
So keep to the maximal number of allowed $\varphi$
functions, at least not reduce the number unnecessarily locally, we
should require that the restrictions of $\varphi$ to $B\cap D$ for any
pair of macro restrictions $\varphi\mid_{\rm restriction ~to ~B \cap D}$ be
the same set of functions on $B\cap D$ from both $(R_D,D)$ and
$(R_B,B)$.  Therefore we must require
\begin{eqnarray}\label{4.2}
\{\hat{\varphi}:B\cap D\to F\mid\hat{\varphi} {\rm ~extendable ~to
~belong ~to~} R_B {\rm ~on} ~B\} \nonumber \\
=\{\hat{\varphi}:B\cap D\to F\mid\hat{\varphi} {\rm ~extendable ~to
~belong ~to~} R_D {\rm ~on} ~D\} 
\end{eqnarray}

We can say that this requirement ensures that the macro restriction
$(R_B,B)$ does not - alone at least - require any drastic selection
of the possible solutions of $\varphi$ to $D$ which are to be in
$R_D$.  If this condition (\ref{4.2}) is not satisfied then
e.g. most elements (ordered sets) in the set of $R_D$ restricted to
$B\cap D$ are not extendable into functions over $B$ that belong to the set
of functions $R_B$.  
But that means that the ones which \underline{can}
be extended into $R_B$ functions are only 
a tiny subset of all the functions in $R_D$, so that if an $R_B$ function
is realized it would seem miraculous.

We would like to put in a remark about the very low accuracy with which
we really intend to work here:  Since our main goal is to obtain
results about entropy flowing and conservation and entropy in an in
most situations to be considered tiny Boltzmann's constant $k$ as unit.
We will be satisfied with an accuracy in the number of $\varphi$ function
(by solving the restrictions).
It is so low that deviations by a few orders of magnitude is considered
quite negligible difference.

In the same spirit we also take it that two different sets of $\varphi$ functions 
have typically numbers of elements which deviate by several orders of magnitude.
Thus one will generically dominate the other one.
In fact entropy is $k$ times the logarithm of the number of micro states
in a macro state and thus we do not need a high accuracy.  So if the number
of elements in the two sets do not differ by a big factor it
will not be so significant.  
Therefore suppose that there is
indeed a significant lack of matching of the two sets
of functions on $B\cap D$ differ in number by such a big factor.
Then if you find a $\varphi\mid_{\rm restriction ~to ~D}$  which matches
$(R_B, B)$ by having its restriction to $B \cap D$ extendable into an
$R_B$ function then it would seem very unlikely.

It should be recognized that our assumption (\ref{4.2}) actually imply that 
we have only reversible processes and that the second law of thermodynamics is also
fulfilled in the trivial manner $\dot{S}=0$.
In fact  we could imagine that we attempted to have a system of macro states and
macro states such that entropy increased (i.e. irreversible process).
In this case we know that at later times $t_l$ than the era of the irreversible
process the possible micro developments -- in our notation $\varphi$
is a field in space time --
could not be constructed.
It represents all the configurations allowed by the high entropy macro states at such
a later time $t_l$.
In other words there would be many states in the high entropy macro state at late time,
but they could not be realized via equations of motion from a micro state in  a 
previous low entropy macro state.

We may now argue that somewhere in between the two times, ``low" and ``high" entropies
as discussed above, there must be some moment and somewhere, where we have some overlap
between two sets of lattice points associated with equations of motion restrictions
in which (\ref{4.2}) is violated.
We must have some moment of time from which we start to get those macro restrictions.
There all the allowed micro states cannot be realized, because of the relation
by equation of motion to a moment shortly before.
If we denote by $(R_B, B)$ the earliest macro state restriction while the offensive restriction
by $(R_D, D)$, we might reach a situation illustrated in 
Figure 3.
\begin{figure}[htbp]
  \begin{center}
    \includegraphics[keepaspectratio=true,height=35mm]{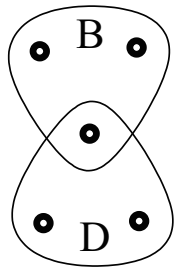}
  \end{center}
  \caption{Schematical figure of overlapping of two subsets B and D on the lattice.}
    \label{fig:e30.eps}
\end{figure} 
In order that the equation of motion restriction symbolized by $(R_D, D)$ or 
rather Fig. 3 will disturb 
all micro state possibilities in $(R_B, B)$, there must be a non-empty overlap 
$B \cap D\neq \phi $.
Since we actually selected the $(R_B, B)$ and $(R_D, D)$ such that the imposition of 
$\varphi$ obeying $R_D$ should diminish the development $\varphi$ that is allowed 
for the macro restriction $(R_B, B)$
we will not fulfill (\ref{4.2}).
Rather we should have 
\begin{eqnarray}
\left\{
\varphi \mid_{B \cap D} \Big|
\varphi \mid_{B} \in R_{B} \wedge \varphi_{D} \in R_{D}
\right\}\nonumber\\
\subset 
\left\{
\varphi \mid_{B \cap D} \Big|
\varphi \mid_{D} \in R_{D}. 
\right\}
\end{eqnarray}
Thus we see that, if we assume validity of (\ref{4.2}) in all the cases of overlapping
restrictions which are either macro -- or equation of motion ones,
then there is no place for irreversible processes.

By making the similar argument time reversed way we can also find that our assumption
(\ref{4.2}) also means that the entropy cannot decrease because that would imply that
at some earlier than the entropy decreasing era the micro states allowed by the 
macro restriction could not be realized due to equations of motion.

This completes the argument that our assumption (\ref{4.2}) leads to be satisfied trivially
second law so that no irreversible processes occur.

\section{An example in one dimension}

To illustrate our formalism we may use it to a simple case, 
one dimensional lattice of
time moments and a general mechanical system developing through a
series of macro states $A(t)$ with entropies $S(A(t))$, at a site $t$.

At first this models has the following two types of macro restrictions:\\
1) The ones specifying just the restriction of $\varphi$ to a single
   discrete site $t$ by
\begin{equation}
\varphi\mid_{{\rm restriction ~to}~\{t\}}
\in R_t=\{\hat{\varphi}:\{t\}\to F\mid\varphi(t)\in A(t)\}
\end{equation}
2) The ones restricting the restriction of $\varphi$ to two
   neighboring sites on $t$-axis i.e. to $\{A,A+a\}$ say.  These are
   the macro restrictions implementing the equations of motion and
   have the form
\begin{eqnarray}\label{5.2}
R_{\left\{t, t+a\right\}}&=&
\left\{
\varphi\mid_{{\rm restriction ~to}~\{t,t+a\}}\in R_{\{t,t+a\}} 
\right\}\nonumber \\
&=&\left\{\hat{\varphi}:\{t,t+a\}\to F\Big|
\hat{\varphi}(t+a)
=\hat{\varphi}(t)+a \eta\frac{\partial
H(\hat{\varphi}(t))}{\partial\hat{\varphi}}\right\}
\end{eqnarray}
which is nothing but the discretized Hamilton equations.
Here $\eta$ is the antisymmetric matrix which takes the form
\begin{eqnarray}
\eta=
\left(
\begin{array}{cc}
0 & \underline{1}\\
\underline{-1} & 0
\end{array}
\right).
\end{eqnarray}
with an appropriate
ordering of the components of $\hat{\varphi}$ conceived of as
generalized coordinates and their conjugate momenta.
We argued above that one should not consider macro restrictions
for subsets of lattice subsets used for other macro restrictions.
Rather one should combine them into macro restrictions for the biggest
of the subsets so as to end up with subsets either fully disjoint or
only partly overlapping.

In the example here it means that we shall so to speak absorb the
macro restrictions for $\{t\}$ and $\{t+a\}$ into that for
$\{t,t+a\}$.  In this way we get
replacement macro restrictions for the subsets of the lattice of the
form $\{t,t+a\}$.  In fact we get
the following simple expression, 
\begin{eqnarray}\label{5.4}&&
\varphi \mid_{{\rm restriction ~to} ~\{t,t+a\}} \in
R^{repl.}_{\{t,t+a\}} \nonumber \\&&
{\tiny
 =  \Big\{\hat{\varphi}:\{t,t+a\} 
 \to  F \Big|
\hat{\varphi} (t) \in A(t) \wedge \varphi (t+a) \in A(t+a) \wedge
\hat{\varphi} (t+a) 
 =  \hat{\varphi} (t)+a \eta \frac{\partial
H(\hat{\varphi}(t))}{\partial\hat{\varphi}} \Big\}
}
\nonumber\\
\end{eqnarray}

We either work with undiscretized phase space for $F$ or we
discretize into points with a given density so that each point
gets a cell of volume $h^N$ where $N$ is the number of degrees of
freedom, $n=2N$. It is a density proportional to phase
space density.  In any case we can use that the time development under
the Hamilton equations is a canonical transformation and as such
conserves the phase space volume.  This means that the original
relation $R_{\{t,t+a\}}$ from (\ref{5.2}) allows just one
value at $t$ for each value at $t+a$ and oppositely.  If the two entropies
$S(A(t))$ and $S(A(t+a))$ happen to be equally big it is at least
possible that the replacing macro restriction (\ref{5.4}) simply allows
realization of any point in $A(t)$ as well as any point in $A(t+a)$.
However if these two entropies are not equal it will be the smaller
one of the two entropies $S(A(t))$ and $S(A(t+a))$ that determines the
size of the set $R^{repl.}_{\{t,t+a\}}$ of restrictions of $\varphi$
to $\{t,t+a\}$ allowed by the combined macro restriction eq.(\ref{5.4}).  In
fact
\begin{equation}
\log Cn R^{repl.}_{\{t,t+a\}}=\min \{S(A(t)),S(A(t+a)) \}.
\end{equation}
If we go to consider if two neighboring ones out of these
replacing macro restrictions $R_{\{t,t+a\}}$ and $R_{\{t-a,t\}}$ 
we ask whether the restrictions to the overlap region
$\{t-a,t\}\cap\{t,t+a \}=\{t\}$ can match.  For such a matching it
would at least be needed that the cardinal number $Cn$ for the two
different sets of functions would be the same:
\begin{eqnarray}
&&
Cn \left\{ \hat{\varphi}:\{ t\} \to F \Big|\hat{\varphi} {\rm ~extendable
~to} ~R^{repl.}_{\{t,t+a\}}\right\}  \nonumber \\&&
=Cn \left\{\hat{\varphi}:\{ t \} 
\to F \Big|\hat{\varphi} {\rm ~extendable  
~to ~an} ~R^{repl.}_{\{t-a,t\}} {\rm function}\right\}.
\end{eqnarray}
But now we quickly see that these numbers are 
$\min\{S(A(t)),S(A(t+a)\}$ and $\min\{S(A(t-a)),S(A(t))\}$ respectively.
The condition for any chance of matching is therefore
needed that
\begin{equation}\label{5.7}
\min\{S(A(t)),S(A(t+a)\}=\min\{S(A(t)),S(A(t-a)\}.
\end{equation}

We can also say that the only entropy numbers relevant for the
replacement  $R^{repl.}_{\{t,t+a\}}$'s, namely $\log Cn
R^{repl.}_{\{t,t+a\}}$ have to be the same i.e. $\log Cn
R^{repl.}_{\{t,t+a\}}=\log Cn R^{repl.}_{\{t-a,t\}}$.  That is
to say for all the information about the entropies of the $A(t)$'s it
has left any footprint into the $\log Cn R^{repl.}_{\{t,t+a\}}$ and
actually thereby to the physics of the model these entropies $S(A(t))$
must be constant down along the chain of time.  We could simply use
the Taylor expansion of a smooth $S(A(t))$ as function of $t$ and insert
it into our condition eq.(\ref{5.7}) and we would deduce
\begin{equation}
\dot{S}(A(t))=0.
\end{equation}
In other words, our matching condition ``to have no miracles locally''
(in either way of time direction thinking) leads to the constancy of
entropy in the one dimensional example which we just presented.

\section{The two dimensional example}

Using the same procedure in our two dimensional lattice example we
take it that the biggest subsets of the lattice for which we have
macro restrictions are the three element ones associated with the
equation of motion.  Basically we should then absorb the other
macro restrictions into these three point ones, which should then
replace all the others.  After this replacement - a kind of absorption -
there will be no longer any tracks left of macro restrictions
associated with subsets $B$ of the lattice contained in the surviving
three-point subsets.  Suppose e.g. that the macro states imposed on
the three sites in a subset $D$ of the lattice associated with the
equation of motion are called $G$, $J$, $Q$.  That is to say we have
in the lattice depicted in Fig. 4:
\begin{center}
 \includegraphics[width=6cm]{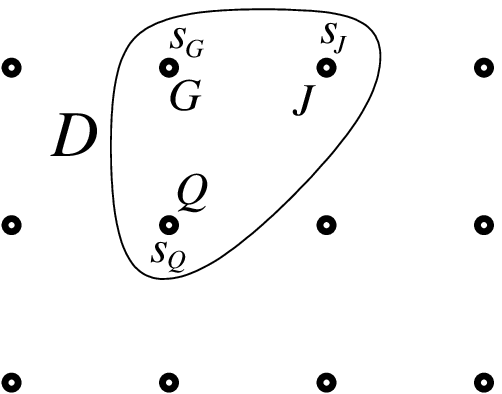}\\
\begin{figure}[h]
  \caption{Equation of motion for three points on the lattice}
\end{figure}
\end{center}
Here $s_G$, $s_J$ and $s_Q$ are the names of the sites.
Whatever these macro restrictions to the macro states $G$, $J$, and
$Q$ originally would have been the information surviving into the
replacement $R^{repl.}_D$ macro restriction from these macro states
would only be some macro states which could then be called $G_D$, $J_D$ and
$Q_D$.  They will be defined from the $R^{repl.}_D$ as
\begin{eqnarray}
G_D &\hat{=}& \{ \hat{\varphi} (s_G) \mid \hat{\varphi}: \{s_G,s_J,s_Q\}
\to F \wedge \hat{\varphi} \in R^{repl.}_D\} , \nonumber \\
J_D &\hat{=}&\{ \hat{\varphi}(s_J)\mid\hat{\varphi}:D\to F \wedge
\hat{\varphi}\in R^{repl.}_D\} ,\nonumber \\
Q_D &\hat{=}& \{ \hat{\varphi}(s_Q)\mid\hat{\varphi}:D\to F \wedge
\hat{\varphi}\in R^{repl.}_D\}. \nonumber \\
\mbox{where}~~~
\{s_G, &s_J,& s_Q\}=D.
\end{eqnarray}
Defined from another set of three points $E$ say which also contains say
$s_G$ we get analogously
\begin{equation}
G_E\hat{=}\{\hat{\varphi}(s_G)\mid\hat{\varphi}:E\to F \wedge
\hat{\varphi}\in R^{repl.}_E\}.
\end{equation}
The condition to avoid locally miraculous restrictions - our principle
- now comes to say e.g.
\begin{equation}
G_D=G_E
\end{equation}
for a situation like:
\begin{center}
  \includegraphics[width=8cm]{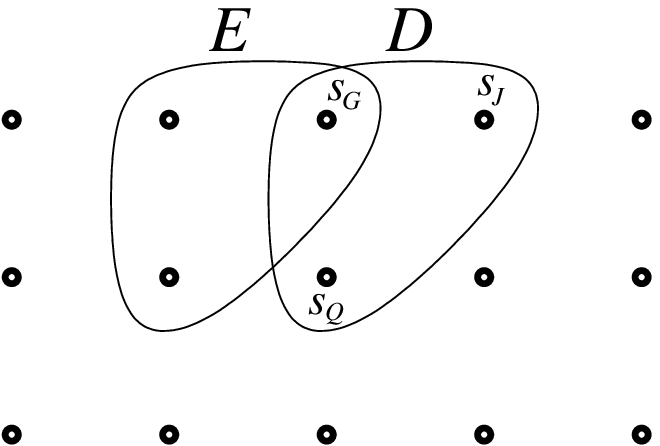}\\
\begin{figure}[h]
  \caption{Equations of motion leading to $G_D=G_E$}
\end{figure}
\end{center}  
This relation implies $S(G_D)=S(G_E)$ for the
entropies.

In the spirit of only the information surviving into the
$R^{repl.}_E$, $R^{repl.}_D$, etc accessible after all at all we
could also redefine correlation of entropies between neighboring points -
whenever present in one of our three - point $B$-sets - by putting say
\begin{eqnarray}
S(G_D)+S(J_D)+(M_{GJ})_D
=\log Cn \{ \hat{\varphi}:\{s_G,s_D\}\to F \mid\hat{\varphi}:D\to F\wedge
\hat{\varphi}\in R^{repl.}_D\}.
\nonumber\\
\end{eqnarray}

\section{Defining three different entropy currents
$j^{\mu}_A$, $j^{\mu}_B$ and $j^{\mu}_C$}

We have found above a sort of conservation rules associated with
parallelogram-like figures.  We should have in mind that there are in 
our lattice \underline{three} such types of parallelogram-like
structure orientations.  It is therefore natural that we shall
construct three different entropy currents, which we may in the
continuum limit denote $j^{\mu}_A$, $j^{\mu}_B$ and $j^{\mu}_C$
corresponding to the three different orientations of parallelograms
to be associated with the conservation.

Before writing down complete form of the continuum limit currents
$j^{\mu}_A$, $j^{\mu}_B$ and $j^{\mu}_C$ by means of the lattice
quantities as it is our goal to do soon, we should look a bit more on
how it comes that we have these three different concepts of entropy
flow.

For this purpose we want to point out corresponding to one of the
orientations of our parallelogram-like structures there is a restricted
class of curves on the lattice.  Such a class of curves is by
definition a curve of the type allowed for either the full drawn or
the broken lines forming what we could call the two sides of the 
``parallelogram-structures'' as is depicted in Fig. 6.  
These curves may be described by
thinking of them as oriented meaning with an arrow along them and then
they are allowed to be composed from only two types of links.  For instance we
define corresponding to a parallelogram-like structure of the
orientation
\begin{center}
 \includegraphics[width=8cm]{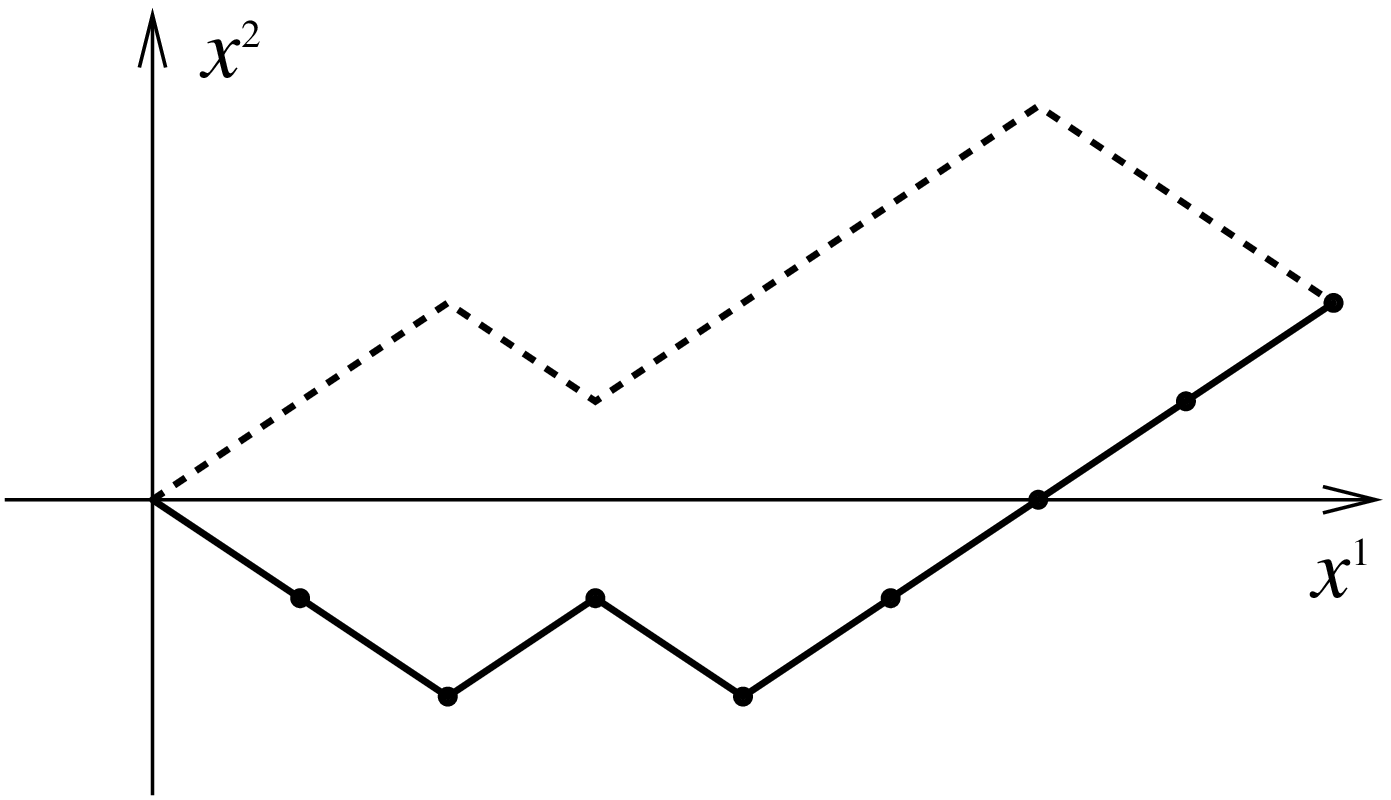}
\begin{figure}[h]
  \caption{Two sides of parallelogram structure}
\end{figure}
\end{center}
have that in the ``positive'' direction along the
curves allowed in the class which we call here class $A$, there are only links
corresponding to space time vectors
($\sqrt{\frac{3}{4}}a$,$\frac{1}{2}a$) and
($\sqrt{\frac{3}{4}}a$,-$\frac{1}{2}a$).  Here $a$ is the hexagonal
lattice constant, i.e. the length of the sides of the triangles.
Starting from a lattice site $x^{\mu}$ say we can by composing these
two vectors in succession construct an infinite number of half-curves
extending from this point $x^{\mu}$ as in figure 7:
\begin{center}
 \includegraphics[width=12cm]{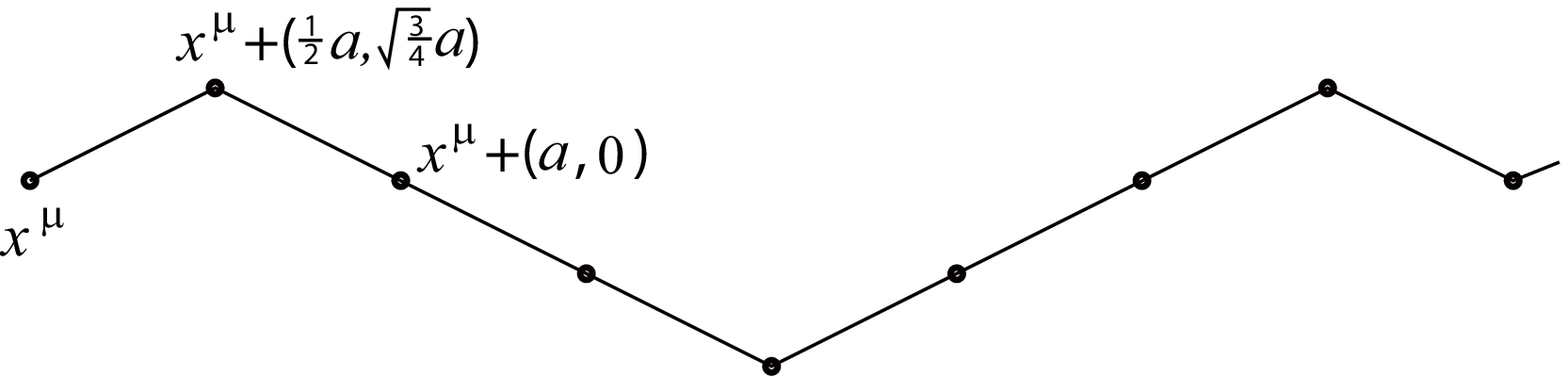}
\begin{figure}[h]
  \caption{Half-curve with only the positive direction}
\end{figure}
\end{center}
We will call such a family of half-curves extending from $x^{\mu}$ the
$A$-class of half curves extending from $x^{\mu}$.  It will be later of
interest that with the equations of motion associated to every other
of the triangles in the lattice as already described, we can use these
equations of motion to predict the restriction of a solution to the
equations of motion to one of the half-curves in class $A$ extending
from $x^{\mu}$ say $\alpha$ to the restriction of an other half curve in the
same class $A$ also extending from $x^{\mu}$, say $\beta$.  
\newcommand{\LINE}{{\raisebox{-0.0em}{\includegraphics[width=1.0cm]{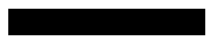}}}}
\newcommand{\TRIANGLE}{{\raisebox{-0.05em}{\includegraphics[width=0.4cm]{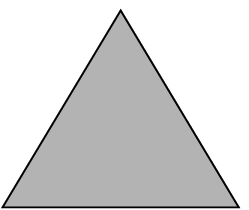}}}}
\newcommand{\TRIANGLEB}{{\raisebox{-0.2em}{\includegraphics[width=0.4cm]{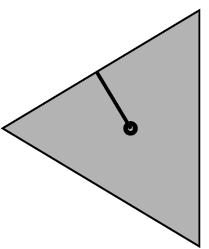}}}}
\begin{center}
 \includegraphics[width=13cm]{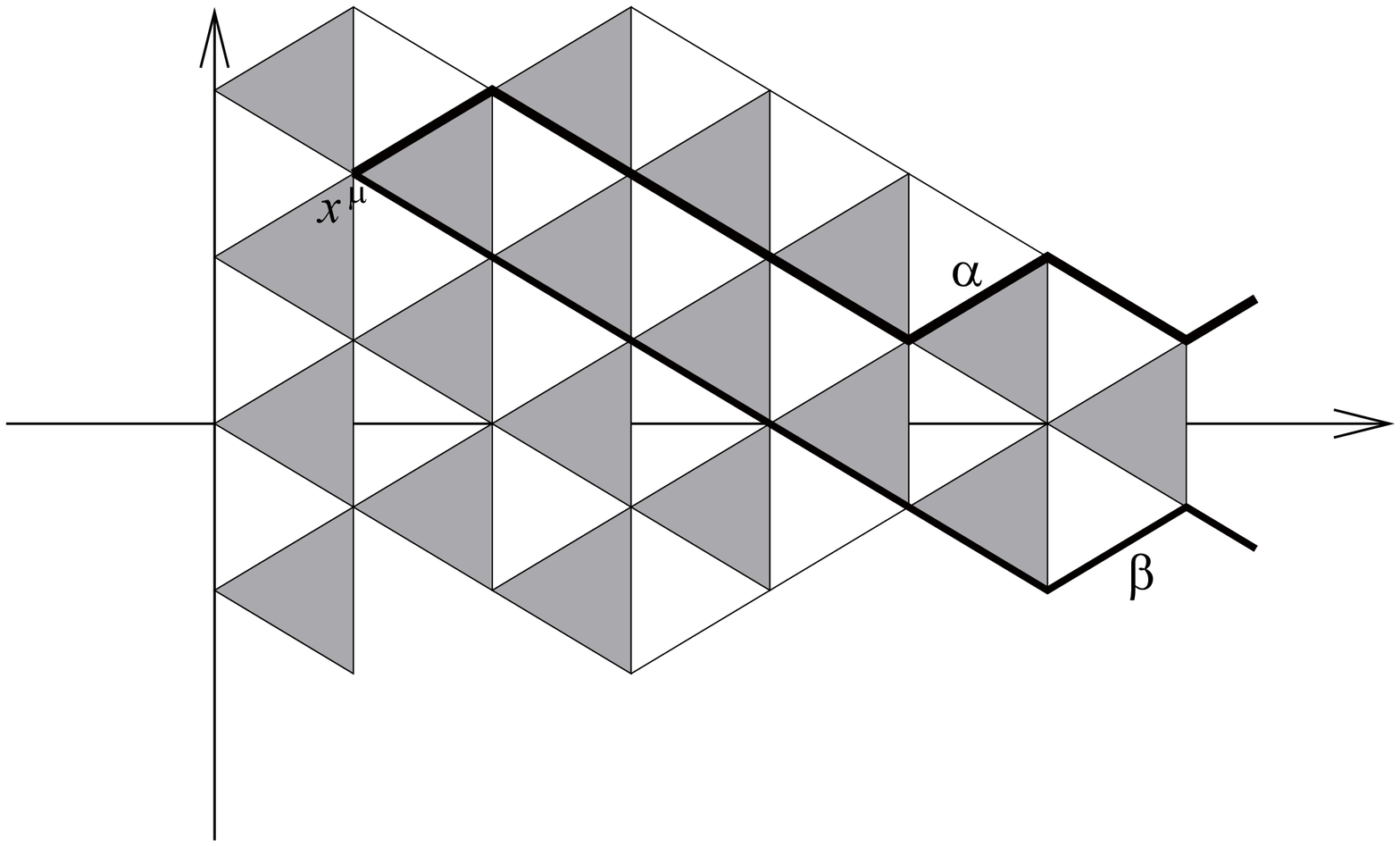}
\begin{figure}[h]
  \caption{Two half-curves in class A, both of which start from $x^\mu$.
  Two points of the black triangle 
  uniquely specifies the 3rd point and its field values 
  which give a restriction to solutions of equations of motion.}
\end{figure}
\end{center}
In fact one may
easily convince oneself by looking at figure 8
that supposing a solution on the half curve $\alpha $ - here $\alpha $ is written
with the line \LINE - is given, then the values of that solution on the
$\beta $ - half-curve can be calculated without further input information.
In fact one shall just use the rule that when at two points in a black
triangle \TRIANGLE\ the solution values are known then the value on the
third corner of this triangle is uniquely given by equation of
motion.  Basically the uniqueness is argued for it relatively from
curve $\alpha $ to curve $\beta $.  
From this unique predictability from one half-curve $\alpha $
to another one $\beta $ in the class
we may think of a speculation of $\varphi$  along any curve in class $A$
from $x^{\mu}$ to tell the same information about $\varphi$.
That
is to say that by choosing say the class $A$ one has specified in
detail what could be called the ``information'' about a solution
$\varphi$ to the equations lying to the $A$-side of a space-time point $x^{\mu}$.  
We will say that the information about solutions
$\varphi$ contained to the $A$-side of $x^{\mu}$ is the information
contained in the restriction of $\varphi$ to one of the half-curves,
$\beta $ say, extending from $x^{\mu}$ and belonging to class $A$, which
again means composed successively from
($\frac{1}{2}a$, $\sqrt{\frac{3}{4}}a$) and
($-\frac{1}{2}a$, $\sqrt{\frac{3}{4}}a$).  

As stressed it does 
not matter which of these half-curves we use, they all give the same
information about solutions.  This question of information about
solutions really gets precise by talking about that
solutions $\varphi$ can be classified into classes inside which the
solutions have the same restriction to all the half-curves of $A$-type
extending from $x^{\mu}$.  
But again we have to have in mind it is enough to
require two solutions $\varphi_1$ and $\varphi_2$ to have the same
restriction
\begin{equation}
\varphi_1\mid_\alpha =\varphi_2\mid_\alpha 
\end{equation}
to one half-curve in the class of curves then it will be true for all
the half-curves of type $A$ extending from $x^{\mu}$.

If you therefore want to count the number of allowed solutions
corresponding to a macro scenario and to define how the information
needed to specify a solution lies to one side or to the other side of
the space time point $x^{\mu}$, we may make the concept of these
``sides'' meaningful by saying that the information on the $A$-side is
the one contained in the restriction of $\varphi$ to one of the
half-curves, $\alpha $, i.e. the information contained in
$\varphi \mid_\alpha $.

Ignoring the problems of infrared divergences we could say that, if we
consider the set of all solutions $\{\varphi\}$ of a given macro
scenario and ask for how many different
restrictions $\varphi$$\mid_\alpha $ to a half-curve $\alpha $ in the $A$-class
extending from $x^{\mu}$, the logarithm of this number represents the
entropy to the $A$-side of $x^{\mu}$.  
The counting just by numbers is
only possible if we have discretized the value space
for solutions $\varphi$ so that they become countable, but then one
could instead imagine using a method of the type of measuring phase
space volume.  Even if infrared divergence problems makes it
ill-defined what the full amount of entropy to the $A$-side of $x^\mu$
is, we can still get quite meaningful convergent results for the
entropy on the $A$-side of $x^\mu$ subtracted from that to the
$A$-side of another space time point $y^\mu$, because we can now find
$A$-type half courses extending from $x^\mu$ going to follow or coincide
with an $A$-type half-curve extending from $y^\mu$ from some point on
as depicted in Fig. 9:
\begin{center}
 \includegraphics[width=11cm]{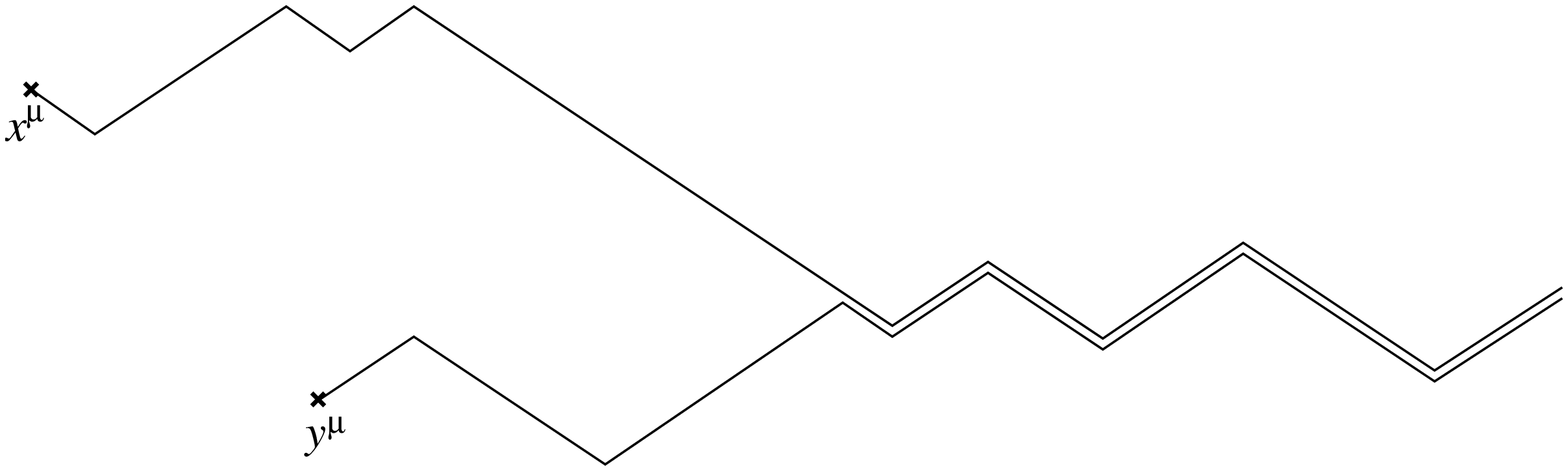}
\begin{figure}[h]
  \caption{The $A$-type half-curve of $x^\mu$ coincides with an another
  $A$-type half-curve extending from $y^\mu$}
\end{figure}
\end{center}  

Then we can use alone the restrictions of the solutions to the
pieces of the half-curve which are not coinciding, and these pieces
will be finite.

It should be kept in mind that we want to identify the
entropy on the $A$-side of $x^\mu$ with the number of solutions in the
given macro scenario which is characterized by each class having its own
restriction to the type A half-curves extending from $x^\mu$.

It should thus be understood that to the extend that we can identify
this newly defined concept to the $A$-side of a space time point
$x^\mu$ from a more usual concept of one side of a point in a one-space
dimension world we may use such a concept to make a meaning where
entropy is placed and thus entropy density and furthermore entropy flow.

\section{Equations for $S(x)$ and $\log Cn R(x)$}

We want to compute what the conservation laws
$\partial_\mu$$j^\mu_A$= $\partial_\mu$$j^\mu_B$=
$\partial_\mu$$j^\mu_C$=0 for the three currents mean for the two
scalar fields $S(x)$ and $\log Cn R(x)$ in terms of which we managed
to write them.  Since $S(x)$ always come into the expressions for the
currents via the difference $S(x)- \log Cn R(x)$ we shall give this
quantity the name $D(x)$
\begin{equation}
D(x)\hat{=}S(x)-\log Cn R(x).
\end{equation}
Let us also define for the three types half-curve systems $A$, $B$,
and $C$ the unit vectors orthogonal to the average direction of these
half-curves
\begin{equation}
\theta^\mu_A=(0,1)^\mu,\hspace{5mm}
\theta^\mu_B=(-\sqrt{\frac{3}{4}},-\frac{1}{2})^\mu, {\rm and}
~\theta^\mu_C=(-\sqrt{\frac{3}{4}},-\frac{1}{2})^\mu
\end{equation}
as well as the vectors in these average directions
\begin{equation}
H^\mu_A=(1,0)^\mu,\hspace{5mm}
H^\mu_B=(-\frac{1}{2},\sqrt{\frac{3}{4}})^\mu, {\rm and}
~H^\mu_C=(-\frac{1}{2},-\sqrt{\frac{3}{4}})^\mu.
\end{equation}
Notice that you obtain $\theta^\mu_A$ by rotating $H^\mu_A$ by
$90^\circ$ anticlockwise and analogously $\theta^\mu_B$ by rotating
$90^\circ$ $H^\mu_B$ and so on.  Using this notation we
can write expressions for the three entropy currents become
\begin{eqnarray}
j^\mu_A(x)&=&\theta^\mu_A 
\left\{D(x)+\frac{a}{\sqrt3}H^{\rho}_A\frac{\partial}{\partial{x^{\rho}}}\log
CnR(x)\right\}+\frac{\sqrt3{a}}{4}\varepsilon^{\mu\nu}\frac{\partial}{\partial{x^\nu}}\log
Cn R(x) \nonumber \\
j^\mu_B(x)&=&\theta^\mu_B 
\left\{D(x)+\frac{a}{\sqrt3}H^{\rho}_B\frac{\partial}{\partial{x^{\rho}}}\log
CnR(x)\right\}+\frac{\sqrt3{a}}{4}\varepsilon^{\mu\nu}\frac{\partial}{\partial{x^\nu}}\log
Cn R(x) \nonumber \\
j^\mu_C(x)&=&\theta^\mu_C 
\left\{D(x)+\frac{a}{\sqrt3}H^{\rho}_C\frac{\partial}{\partial{x^{\rho}}}\log
CnR(x)\right\}+\frac{\sqrt3{a}}{4}\varepsilon^{\mu\nu}\frac{\partial}{\partial{x^\nu}}\log
Cn R(x).
\end{eqnarray}
Since the topological current term
$\frac{\sqrt3{a}}{4}\varepsilon^{\mu\nu}\frac{\partial}{\partial{x^\nu}}\log Cn R(x)$ 
is trivially conserved the nontrivial information from the
conservation of the three currents comes only from the conservation of
the first parts and that tells
\begin{eqnarray}
\theta^\mu_A\partial_\mu D+\frac{a}{\sqrt3}\theta^\mu_A
H^{\rho}_A\partial_\mu\partial_\rho\log Cn R(x)&=&0 \nonumber \\
\theta^\mu_B\partial_\mu D+\frac{a}{\sqrt3}\theta^\mu_B
H^{\rho}_B\partial_\mu\partial_\rho\log Cn R(x)&=&0 \nonumber\\
\theta^\mu_C\partial_\mu D+\frac{a}{\sqrt3}\theta^\mu_C
H^{\rho}_C\partial_\mu\partial_\rho\log Cn R(x)&=&0.
\end{eqnarray}

It is easy to see - as we essentially already did - that the sum of
these three equations is trivially zero just from
$\partial_p$$\partial\mu$=$\partial\mu$$\partial_p$.  So there is in
reality only two independent equations, just enough to determine
$D(x)$ up to an additive constant.  
However, one may wonder whether
there exists any $D(x)$ function at all satisfying these equations.
Indeed the condition for that to be the case is that the partial
derivatives of second order for $D(x)$ derived from these equations
are consistent with commutativity of the partial derivatives.  For
that we can consider the condition that
\begin{equation}\label{8.6}
(\theta^\mu_A\theta^\nu_B-\theta^\mu_B\theta^\nu_A)\partial_\mu\partial_\nu
D=0
\end{equation}
The idea to obtain eq.(\ref{8.6}) is to 
differentiate the first of the three equations in the
$\theta^\mu_B$-direction, i.e. act on it with
$\theta^\mu_B$$\partial_\mu$, and then compare with what we get from
the second equation by acting with $\theta^\mu_A$$\partial_\mu$.  It
is easily seen the consistency condition for the two first equations
manipulated this way becomes
\begin{equation}
(H^\mu_A-H^\mu_B)\theta^\nu_A\theta^\rho_B\partial_\mu\partial_\nu\partial_\rho
\log Cn R(x)=0.
\end{equation}
Since indeed $H^\mu_A-H^\mu_B\propto\theta^\mu_C$ and we 
can symmetrize in the three indices $\mu$, $\nu$, $\rho$
because $\partial_\mu\partial_\nu\partial_\rho\log Cn R(x)$
is a symmetric third rank tensor we may write this condition
\begin{equation}
\theta^{(\mu}_A\theta^\nu_B\theta^{\rho)}_C\partial_\mu
\partial_\nu\partial_\rho\log Cn R(x)=0.
\end{equation}
We can consider the symmetric third rank tensor
\begin{equation}
\hat{\xi}^{\mu\nu\rho}=\theta^{(\mu}_A \theta^\nu_B \theta^{\rho)}_C
\end{equation}
connected with our lattice and the selection of
triangles to be associated with equations of motion.

It is easy to see that if indeed $\log Cn R(x)$ obey the third order
homogeneous differential equation then we can from $\log Cn R(x)$
integrate up to an additive constant.

\section{The stressing of the three types of entropy current}

Now, however, it is crucial for our discussion that there is on our
lattice three types of entropy currents under rotation by $120^\circ$ in the
space-time plane choices of half-curve classes or equivalently
parallelogram-like structure of orientations.  We denote them by $A$,
$B$, and $C$ and we have already discussed that the curves in class
$A$ are made by successively compassing steps
$(\sqrt{\frac{3}{4}}a,\frac{1}{2}a)$ and
$(\sqrt{\frac{3}{4}}a,-\frac{1}{2}a)$.  Analogously we let the
$B$-type of half-curves be compared successively from steps 
$(0,a)$ and $(-\sqrt{\frac{3}{4}}a,\frac{1}{2}a)$.  
And finally we have also a class of half-curves defined by the steps $(0,-a)$, and
(-$\sqrt{\frac{3}{4}}$a,-$\frac{1}{2}$a).  It should be seen
that apart from the sign with which the links are allowed into the
three defined types of half-curves which are thought of as the one
oriented away from the starting point $x^\mu$, there are only three
directions of links in the lattice that can be used as steps for the
curves.  Thus for example  $(\sqrt{\frac{3}{4}}a,-\frac{1}{2}a)$ used
in the $A$-type half-curves is re-used with opposite sign,
$(-\sqrt{\frac{3}{4}}a,\frac{1}{2}a)$ as a step for type $B$.

Let us also have in mind that there is one type of what we called
parallelogram-like structures that can be called type $B$, one that can
be called type $A$ and so on.

\section{Construction of the three entropy currents}

\newcommand{\oo}{{\raisebox{-0.0em}{\includegraphics[width=0.3cm]{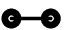}}}}
\newcommand{\oO}{{\raisebox{-0.0em}{\includegraphics[width=0.3cm]{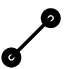}}}}
\newcommand{\Oo}{{\raisebox{-0.0em}{\includegraphics[width=0.3cm]{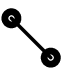}}}}
\newcommand{\oA}{{\raisebox{-0.0em}{\includegraphics[width=0.3cm]{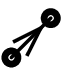}}}}
\newcommand{\Oa}{{\raisebox{-0.0em}{\includegraphics[width=0.3cm]{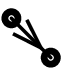}}}}
\newcommand{\ov}{{\raisebox{-0.0em}{\includegraphics[width=0.12cm]{ov.eps}}}}

To construct continuum limit entropy flows we may choose one of
the following routes:\\
1) We may take as starting point of the link-entropy conservation rules
   which really says:

If we form a closed curve from two finite pieces of curves of say type
$\alpha $, and $\beta $, - so that they form together one of the parallelogram-like
structures of orientation $A$ - then the sum of the link-entropies is
\begin{equation}
K_{GH}=\frac{1}{2}(S_G+S_H)-\log Cn (R_{GHL})
\end{equation}
where along $\alpha $ and along $\beta $ are equal
\begin{eqnarray}
 \sum_{\oo\in \alpha } K_{\oo} &=& \sum_{\oo\in \beta } K_{\oo}
 \label{psra}
\end{eqnarray}
2) We may take as the starting point the entropy definition by means
   of number of different restriction to a half-curve $\alpha$  say,
   i.e. $\varphi\mid_\alpha $, in the solution set for the given
   macro scenario.

Note that in both cases we have to select one of the types of curves $A$,
$B$ or $C$.  We must therefore a priori expect that can and must define
three different entropy currents, each of which have to be marked by this choice
$A$, $B$ or $C$.  That is to say we shall define in continuum limit 
at first three different entropy currents $j^\mu_A(x)$,
$j^\mu_B(x)$ and $j^\mu_C(x)$, all being conserved in the adiabatic case
\begin{equation}
\partial_\mu j^\mu_A(x)=\partial_\mu j^\mu_B(x)=\partial_\mu
j^\mu_C(x)=0.
\end{equation}
It would be unexpected to find so many entropy currents
being truly different, and we would therefore expect - and indeed
shall find below - that these three entropy currents
$j^\mu_A(x)$, $j^\mu_B(x)$ and $j^\mu_C(x)$ are indeed related,
so that they only deviate in a rather trivial way by constants.

\section{Construction starting from the parallelogram-like structure
conservation rule}

Let us now contemplate how to construct a continuum limit entropy
current density say $j^\mu_A(x)$ so that its conservation is related
to the sum of the link entropy rule 
using curve pieces $\alpha $, $\beta $ of the $A$-type.
From the links of the relevance in the $A$-case, the ones going in
direction 
$(\sqrt\frac{3}{4}a, \frac{1}{2}a)$ or 
$(\sqrt\frac{3}{4}a, -\frac{1}{2}a)$ , we can use the 
$K^{\oo}_{GH}=\frac{1}{2}(S_G+S_H)-\log Cn(R_{GHL})$ link entropies to
construct the current density $j^\mu_A(x)$ in the region around the 
link(s) in question.

From the continuum limit approximation we will take it that when the
lattice constant $a$ is small the value $K_{\oo}$ of the link entropy
varies slowly from one link to the neighboring ones of the same direction.
We do not know however a priori any good reason for that links with
different direction should have approximately the same $K_{\oo}$ even in
close to each other in space-time.

A priori we would set up an expression for $j^\mu_A(x)$ of the form
\begin{equation}
 j^\mu_A(x) =
  \begin{array}{c}
   \mbox{\small average}\\[-0.3em]
   \mbox{\small around}\\[-0.3em]
   x
  \end{array}
  \left(b^\mu K_{\Oo}+c^\mu K_{\oO}\right)
\end{equation}
where $K_{\Oo}$ and $K_{\oO}$ symbolize the two different directions of
links relevant for the curves of type $A$, and $b^\mu$ and
$c^\mu$ are some constant space-time vectors (2-vectors) to be
chosen so as to make $j^\mu_A(x)$ conserved and to satisfy possible
other wishes.
The average around $x$  means that we strictly speaking extract
$j^\mu_A(x)$ over a region so large that the lattice structure is no
longer felt.

In the continuum limit we can think of the two link entropies relevant
for case $A$ as two functions $K_{\oO}(x)$ and $K_{\Oo}(x)$ of the
space-time point $x$.
We may also introduce for instance the unit vectors along these link
directions $e^\mu_{\oA}$ and $e^\mu_{\Oa}$.
Then the conservation law expressed by the parallelograms 
can be written in the continuum limit 
\begin{equation}
e^\mu_{\oA}\partial_\mu K_{\Oo} - e^\mu_{\Oa}\partial_\mu K_{\oO}=0
\end{equation}
This would mean that if we defined a current
\begin{equation}
j^\mu_A(x)=b^\mu K_{\Oo}(x)+c^\mu K_{\oO}(x)
\end{equation}
with
\begin{eqnarray}
b^\mu&=&e^\mu_{\oA}\nonumber\\
c^\mu&=&-e^\mu_{\Oa}\nonumber
\end{eqnarray}
then the conservation rule with the parallelogram would lead to the
conservation of this current.

If we choose the $b^\mu$ and $c^\mu$ in an other ratio or in
other directions the current $j^\mu_A(x)$ will not be conserved.

\section{Continuum limit for the expressions for $K_{GH}$ in terms of
$S(x)$ and $\log Cn(R(x))$}

In the continuum limit we should notice that in the expression for the
link entropy
\begin{equation}
K_{GH}=\frac{1}{2}(S_G+S_H) - \log Cn(R_{GHL})
\end{equation}
the center of the triangle $\triangle GHL$ associated with  an equation
of motion element is not quite at the same positions as lattice sites
associated with the ``entropies at sites'' $S_G$ and $S_H$.
Although  in the very crudest approximation we would just put
$K(x)=S(x) - \log Cn(R(x))$, this is therefore not quite true.
We should rather say if we want to $K_{GH}=K(x)$ the $x$ is middle of
the link $GH$ then clearly $\frac{1}{2}(S_G + S_H)$ will even
including linear terms in a Taylor expansion, i.e. of order of the
lattice constant $a$, be equal to $S(x)$. But the center of the
triangle is displaced and we must rather  take
\begin{equation}
\log Cn(R_{GHL})= \log Cn(R(x))+ \chi^\mu \frac{\partial}{\partial x^\mu}
\log Cn(R(x))
\end{equation}
where $\chi^\mu$ is defined by 
\begin{equation}
\chi^\mu =x^\mu_{center \triangle GHL}- x^\mu_{center GH\cdot}
\end{equation}

The $\chi^\mu$ only depends on the lattice orientation and the link
direction and on which of the triangles that are associated with the
equation of motion.
This latter dependence is in fact crucial for the sign of the 2-vector
$\chi^\mu$ because a link $GH$ lies in the lattice between two triangles,
but it is only one of them that is associated with an equation of
motion element.
If one chooses the wrong triangle one would get the opposite sign for
$\chi^\mu$, but $R_{GHL}$ is associated with a triangle in
correspondence with an element of equation of the motion.

In the $A$-type half-curve system for constructing the $A$-type
entropy current $j^\mu_A(x)$ we use the link directions 
$(\sqrt\frac{3}{4}a, \frac{1}{2}a)$ and 
$(\sqrt\frac{3}{4}a, -\frac{1}{2}a)$ and we imagine we have chosen the
lattice so that the triangles associated with equations of motion
\TRIANGLEB\ 
have one tip pointing just in the negative $x^1$-axis direction.
Tip here means an angle in the triangle and that it points in the
negative $x^1$-direction is supposed to that this ``tip'' has a
coordinate only deviating from that of the center of triangle by
having a smaller $x^1$-coordinate.
Taking into account this chosen orientation of our lattice and of
which triangles are equation of motion associated we easily calculate
that :
1) for the links of direction $(\sqrt\frac{3}{4}a, \frac{1}{2}a)$ the
   $\chi^\mu$ pointing from the middle of the link to the center of the
   neighboring triangle associated with the equation of motion is
\begin{equation}
\chi^\mu=(\frac{1}{4\sqrt3}a, -\frac{1}{4}a)
\end{equation}
2) for the links of the direction $(\sqrt\frac{3}{4}a, -\frac{1}{2}a)$ 
we get
\begin{equation}
\chi^\mu = (\frac{1}{4\sqrt{3}}a, \frac{1}{4}a).
\end{equation}
Let us insert
\begin{equation}\label{45}
K_{\oO}(x)=S(x)-\log Cn (R(x)) +(\frac{1}{4\sqrt{3}}a,
-\frac{1}{4}a)^\mu \frac{\partial}{\partial x^\mu} \log Cn (R(x))
\end{equation}
and
\begin{equation}\label{46}
K_{\Oo}(x)=S(x)-\log Cn (R(x)) +(\frac{1}{4\sqrt{3}}a,\frac{1}{4}a)^\mu 
\frac{\partial}{\partial x^\mu} \log Cn (R(x))
\end{equation}
into 
\begin{equation}
j^\mu_A(x)={}^-(\sqrt\frac{3}{4}, \frac{-1}{2})^\mu
K_{\oO}(x)+(\sqrt\frac{3}{4}, \frac{1}{2})^\mu K_{\Oo}(x)
\end{equation}
We obtain
\begin{eqnarray}
j^\mu_A(x)&=&(\sqrt\frac{3}{4}\cdot\frac{1}{2}a 
\frac{\partial}{\partial x^2} \log Cn\left\{R(x)\right\}, 
S(x)-\log Cn\left\{R(x)\right\}\\\nonumber
&+& \frac{a}{4\sqrt{3}} \frac{\partial}{\partial x^1}
\log Cn\left\{R(x)\right\})^\mu\\\nonumber
&=&(0,1)^\mu \left\{S(x)-\log CnR(x) +(\frac{a}{4\sqrt{3}}+\frac{\sqrt{3}a}{4})
\frac{\partial}{\partial x^1} \log Cn R(x)\right\} +\\\nonumber
&+&\frac{\sqrt{3}}{4}a \epsilon^{\mu\nu}
\frac{\partial}{\partial x^\nu} \log CnR(x)\\\label{49}
&=&(0,1)^\mu\left\{S(x)-\log CnR(x) + \frac{a}{\sqrt{3}}
\frac{\partial}{\partial x^1} \log CnR(x)\right\} +\\\nonumber
&+& \sqrt{\frac{3}{16}}a \epsilon^{\mu\nu} 
\frac{\partial}{\partial x^\nu} \log CnR(x)
\end{eqnarray}

We may interpret the quantity
\begin{equation}\label{10.11}
\log CnR(x)-\frac{a}{\sqrt{3}}\frac{\partial}{\partial x^1} 
\log CnR(x)
\end{equation}
as is the extrapolated value of $\log CnR(x)$ to the corner of
triangle at which the two link directions associated with the
$A$-choice meet 
\begin{center}
 \includegraphics[width=3cm]{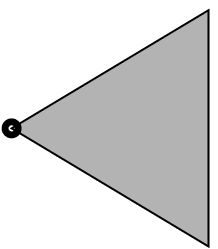}
\begin{figure}[h]
  \caption{The corner of the black triangle at which 
  the two link directions meet, associated with the first term of eq.(\ref{10.11})}
\end{figure}
\end{center}

So if instead of the center of the triangles associated with equations
of motion we had decide to let the representative point of the
triangle be the corner natural in the $A$-case we would have gotten
rid of the extra term 
$-\frac{a}{\sqrt{3}}\frac{\partial}{\partial x^1} \log CnR(x)$
from eq.(\ref{10.11}).
But it should be noted that when we go to the cases $B$ and $C$
another corner would have to be chosen to get rid of the corresponding
terms.

\section{Constructing analogously $j^\mu_B(x)$ and $j^\mu_C(x)$}

For the construction of the three different entropy currents 
$j^\mu_A(x)$, $j^\mu_B(x)$, and $j^\mu_C(x)$ the normalization is 
not given just by ensuring conservation, but even if we normalize
all three numerically in an analogous way there is a sign to be chosen
which is fundamentally arbitrary.
If you think of one direction in space time as the positive time axis
you may say that we shall make the entropy density corresponding to
that have the $K$'s come in with positive coefficients. 
But ignoring second law of thermodynamics as we seek to do in this
article there is no physical difference between positive and negative
time directions.
So such a sign choice for our currents $j^\mu_A(x)$, $j^\mu_B(x)$, and
$j^\mu_C(x)$ is basically arbitrary.
It might even be pedagogical to think of both signs and say that we are
defining six currents $\pm j^\mu_A(x)$, $\pm j^\mu_B(x)$, and
$\pm j^\mu_C(x)$ , rather than only three.

Then analogous construction for the $B$-half-curves associated
$B$-entropy current $j^\mu_B$ goes, since the $B$-choice applies for
the half-curves the links counted with orientation 
\begin{equation}
(0,a), (-\sqrt\frac{3}{4}a, \frac{1}{2}a)
\end{equation}
by using
\begin{equation}
j^\mu_B(x)= (-\sqrt\frac{3}{4}, \frac{1}{2}) K_{\oO}(x) +(0,-1) K_{\Oo}(x)
\end{equation}
where we have chosen a sign corresponding to $j^\mu_B$ as
$j^\mu_A$ rotated by $120^\circ$ in the space time plane.
Then we insert
\begin{eqnarray}
 K_\oO(x)
  &=&
  S(x)
  - \log C_n R(x)
  + \left(-\frac{a}{2\sqrt{3}}, 0\right)^\mu
  \frac{\partial}{\partial x^\mu} \log C_n R(x)
\end{eqnarray}
and
\begin{eqnarray}
 K_\Oo(x)
  &=&
  S(x)
  - \log C_n R(x)
  + \left(\frac{a}{4\sqrt{3}}, \frac{1}{2} a\right)^\mu
  \frac{\partial}{\partial x^\mu} \log C_n R(x)
\end{eqnarray}
and we obtain
\begin{eqnarray}
j^\mu_B
&=&(-\sqrt{\frac{3}{4}}, -\frac{1}{2})
(S(x)-\log Cn R(x))
-(\frac{1}{2\sqrt{3}}a, -\frac{1}{2}a)^\rho 
\frac{\partial}{\partial x^{\rho}} \log Cn R(x))+ \nonumber \\
&& \Bigl\{ (+\sqrt{\frac{3}{4}} \cdot \frac{a}{2\sqrt{3}} 
-\sqrt{\frac{3}{4}} \cdot \frac{a}{2\sqrt{3}}) \partial_{1} \log Cn R(x)
+\frac{1}{2} \sqrt{\frac{3}{4}} a \partial_{2} \log CnR(x), \nonumber \\
&&(-\frac{1}{2} \frac{a}{2\sqrt{3}}-\frac{a}{4\sqrt{3}}-\frac{1}{2}
\cdot \frac{a}{2\sqrt{3}}) \partial_{1} \log Cn R(x)
+(-1\cdot \frac{1}{4}a-\frac{1}{2}(-\frac{1}{2}a))\partial_{2}
\log Cn R(x) \Bigr\} \nonumber \\
&=&(-\sqrt{\frac{3}{4}},-\frac{1}{2}) \left\{ S(x)-\log CnR(x)
-(\frac{a}{2\sqrt{3}},-\frac{a}{2})^{\rho} \frac{\partial}{\partial x^\rho} 
log CnR(x) \right\} \nonumber \\
&&+ \frac{\sqrt{3}a}{4} \varepsilon^{\mu\nu} 
\frac{\partial}{\partial x^\nu} \log Cn R(x)
\end{eqnarray}
Again we can interpret 
$\log Cn R(x)+(\frac{a}{2\sqrt3},-\frac{a}{2})^\rho \frac{\partial}{\partial x^{\rho}} \log Cn R(x)$ 
as the value of $\log Cn R(x)$ if we instead of
identifying the position of the triangle with its center identified it
as the corner selected now by the $B$-choice.

Finally we may now construct the $C$-half-curves related $C-$
entropy current $j^\mu_C(x)$, where we now have the oriented links
from which the half-curves are constructed as
\begin{equation}
(0,-1a), (-\sqrt{\frac{3}{4}}a,-\frac{1}{2}a).
\end{equation}
The $C$-entropy current is with again by $120^\circ$ rotations chosen
sign
\begin{equation}
 j^\mu_C(x)
 = (0,-1) K_\oO(x)
 - \left( -\sqrt{\frac{3}{4}}, -\frac{1}{2}\right) K_\Oo(x).
\end{equation}
Herein we shall insert the already above given expressions for $K_\oO$ and
$K_\Oo$ to obtain
\begin{eqnarray}
j^\mu_C(x)&=&(0,-1)\left\{S(x)-\log Cn
R(x)-(-\frac{a}{4\sqrt{3}},\frac{1}{4}a)^\rho \frac{\partial}{\partial
x^\rho} \log Cn R(x)\right\} \nonumber \\
&+&(+\sqrt{\frac{3}{4}},+\frac{1}{2})\left\{S(x)-\log Cn
R(x)-(\frac{a}{2\sqrt{3}},0) \frac{\partial}{\partial x^\rho} \log Cn
R(x)\right\} \nonumber \\
&=&(+\sqrt{\frac{3}{4}},-\frac{1}{2})\left\{S(x)-\log Cn
R(x)-(\frac{a}{2\sqrt3},\frac{1}{2}a)^\rho \frac{\partial}{\partial
x^\rho} \log Cn R(x)\right\} \nonumber \\
&+&\Bigl\{(-\sqrt{\frac{3}{4}}\cdot
\frac{a}{2\sqrt{3}}+\sqrt{\frac{3}{4}}\cdot 
\frac{a}{2\sqrt3})\partial_1 \log Cn R+(\sqrt{\frac{3}{4}}\cdot
0 \nonumber \\
&+&\sqrt{\frac{3}{4}} \cdot \frac{1}{2}a)\partial_2 \log Cn R,(-1
\cdot(-\frac{1}{4}a)-\frac{1}{2} \cdot \frac{1}{2}a) \partial_2 \log
Cn R \nonumber\\
&+&(-\frac{a}{4\sqrt3}-\frac{a}{2\sqrt3{2}}-\frac{1}{2}\cdot\frac{a}{2\sqrt3}\partial_1  
\log Cn R \Bigr\} \nonumber\\
&=&(\sqrt{\frac{3}{4}},-\frac{1}{2})\left\{S(x)-\log Cn
R(x)-(\frac{a}{2\sqrt3},\frac{1}{2}a)^\rho \partial_\rho \log Cn
R(x)\right\} \nonumber\\
&+&\frac{\sqrt3{a}}{4} \varepsilon^{\mu\nu}
\frac{\partial}{\partial x^\nu} \log Cn R(x)
\end{eqnarray}

\section{Relation between the different entropy currents $j^\mu_A$,
$j^\mu_B$, and $j^\mu_C$}

We may resume by writing the three entropy currents which we have
constructed:
\begin{eqnarray}\label{59}
j^\mu_A(x)&=&(0,1)^\mu\left\{S(x)-\log Cn
R(x)+\frac{a}{\sqrt3}\frac{\partial}{\partial x^1} \log Cn
R(x)\right\}+\frac{\sqrt3{a}}{4} \varepsilon^{\mu\nu}
\frac{\partial}{\partial x^\nu} \log Cn R(x); \nonumber\\
j^\mu_B(x)&=&(-\sqrt{\frac{3}{4}},-\frac{1}{2})^\mu\left\{S(x)-\log Cn
R(x)-(\frac{a}{2\sqrt3},-\frac{a}{2})^\rho \frac{\partial}{\partial
x^\rho} \log Cn R(x)\right\} \nonumber\\
&+&\frac{\sqrt3{a}}{4} \varepsilon^{\mu\nu}
\frac{\partial}{\partial x^\nu} \log Cn R(x); \nonumber\\
j^\mu_C(x)&=&(\sqrt{\frac{3}{4}},-\frac{1}{2})^\mu\left\{S(x)-\log Cn
R(x)-(\frac{a}{2\sqrt3},-\frac{1}{2}a)^\rho \frac{\partial}{\partial
x^\rho} \log Cn R(x)\right\} \nonumber \\
&+&\frac{\sqrt3{a}}{4} \varepsilon^{\mu\nu}
\frac{\partial}{\partial x^\nu} \log Cn R(x).  \nonumber\\
\end{eqnarray}
It is easily seen from these expressions that
\begin{equation}
j^\mu_A(x)+j^\mu_B(x)+j^\mu_C(x)=\frac{a\sqrt3}{4}
\varepsilon^{\mu\nu} \frac{\partial}{\partial x^\nu} \log Cn R(x).
\end{equation}

In fact it is first easily seen that what we could call the main terms
meaning the ones without any $a$-factor cancel
\begin{equation}
\bigl\{(0,1)+(-\frac{\sqrt3}{4},-\frac{1}{2})+(\frac{\sqrt{3}}{4},-\frac{1}{2})
\bigr\}
\cdot \bigl\{S(x)-\log Cn R(x)\bigr\}=0.
\end{equation}

Next we may write the terms connected with shift between the corners
and the center of the triangle in matrix form
\begin{eqnarray}\nonumber
\left\{
\left(
\begin{array}{cc}
0\\
1
\end{array}
\right)
(\frac{a}{\sqrt3},0)+
\left(
\begin{array}{cc}
-\sqrt{\frac{3}{4}}\\
-\frac{1}{2}
\end{array}
\right)
(-\frac{a}{2\sqrt3},\frac{a}{2})+
\left(
\begin{array}{cc}
\sqrt{\frac{3}{4}}\\
-\frac{1}{2}
\end{array}
\right)
(-\frac{a}{2\sqrt3},-\frac{a}{2})
\right\}
\left(
\begin{array}{cc}
\frac{\partial}{\partial x^1} \log Cn R(x)\\
\frac{\partial}{\partial x^2} \log Cn R(x)
\end{array}
\right)
\end{eqnarray}
\[
=
\left(
\begin{array}{cc}
0+\frac{a}{4}-\frac{a}{4} & 0-\frac{a\sqrt3}{4}-\frac{a\sqrt3}{4}\\
\frac{a}{\sqrt3}+\frac{a}{4\sqrt3}+\frac{a}{4\sqrt3} & 0-\frac{a}{4}+\frac{a}{4}
\end{array}
\right)
\left(
\begin{array}{c}
\partial_1 \log Cn R\\
\partial_2 \log Cn R
\end{array}
\right)
\]
\[
=
\left(
\begin{array}{cc}
0 & -\frac{a\sqrt3}{2}\\
\frac{a\sqrt3}{2} & 0
\end{array}
\right)
\left(
\begin{array}{c}
\frac{\partial}{\partial x^1} \log Cn R(x)\\
\frac{\partial}{\partial x^2} \log Cn R(x)
\end{array}
\right)
\]
\begin{equation}
=-\frac{a\sqrt3}{2} \varepsilon^{\mu\nu} \frac{\partial}{\partial
x^\nu} \log Cn R(x)
\end{equation}
together with the three identical topological charge terms
$\frac{\sqrt3{a}}{4} \varepsilon^{\mu\nu} \frac{\partial}{\partial
x^\nu} \log Cn R(x)$ we then obtain
\begin{equation}
(\frac{3\sqrt3{a}}{4}-\frac{a\sqrt3}{2}) \varepsilon^{\mu\nu}
\frac{\partial}{\partial x^\nu} \log Cn R(x)
=\frac{a\sqrt3}{4}\varepsilon^{\mu\nu} \frac{\partial}{\partial x^\nu}
\log Cn R(x).
\end{equation}

\section{Understanding our entropy currents from counting of number of solution}

In this section we should  seek to describe the three above defined
entropy currents $j^\mu_A$, $j^\mu_B$ and $j^\mu_C$ by counting
solutions.
We have already above explained that to each of the three choices
$A$, $B$ or $C$ there corresponds a series of half-curves.
Theses bunches of half-curves have the property that whether two
solutions $\varphi_1$ and $\varphi_2$ to the equations of
motion - obeying also the macro scenario restrictions imposed -
have the same restriction to one of the half-curves, $\delta$ say, is
equivalent to whether they have it on any other, $\gamma$ say, in the same
class of half-curves.
Imagining without going in details that we have chosen both an
infrared cut off and a discretization in a field value space we can
simply ask for the number of classes of solutions $\varphi$ with the
same restriction to a curve in say the $B$-bundle $\delta$, 
$Cn(\{\varphi |_\delta \})$.
We can construct from this number
\begin{equation}
Cn\{\varphi|_\delta|\varphi~~ \mbox{a solution obeying the macro secnario} \}
\end{equation}
what we could call the entropy to the $B$-side of the point 
$x^\mu$, where $\delta$ is taken to extend from $x^\mu$. 
In fact
\begin{equation}
S(B- \mbox{side of}~x^\mu)=\log Cn\{\varphi |_\delta| \varphi~~ \mbox{a solution obeying the macro senario} \}.
\end{equation}

Having defined such an entropy on ``one side'' the current should be
obtained simply as a topological charge out of this 
$S(B-{\rm ~side~of}~x^\mu)$:
\begin{equation}
j^\mu_B(x) = \epsilon^{\mu\nu}\frac{\partial}{\partial x^\nu}
S(B-{\rm ~side~of~}x^\rho)
\end{equation}

Thinking of $x^\rho$ as a site in our triangular lattice and noting
that our link entropies $K_{GH}$ (entropy of link between G and H), if they are 
exponentiated, provide the number of variation possibilities for a
solution restriction to a half-curve $\delta$, say, which comes extra
by having the link $GH$ included.
We can thus see that - even with respect to normalization - the entropy to
the $A$-side of $x^\mu$ can be written as 
\begin{equation}
S(A-\mbox{side of}~ x^\mu) = \Sigma_{{\rm ~along~half-curve}~b} K_{GH}.
\end{equation}
To avoid the infrared divergencies the easiest is to ask for the
difference like
\begin{equation}\label{68}
S(A-\mbox{side of}~ x^\mu)- S(A-\mbox{side of}~ y^\mu) =
\Sigma _{\mbox{along a $B$-type curve for}~ x^\mu ~\mbox{to}~ y^\mu} K_{GH}.
\label{gthrs61}
\end{equation}
Now the links in the $A$-type half-curves go in the directions
$(\frac{3}{4}a, \frac{1}{2}a)$ or $(\frac{3}{4}a, -\frac{1}{2}a)$
and the links of these two directions are displaced by
$\chi^\mu = (\frac{1}{4\sqrt{3}}a, -\frac{1}{4}a)$ and 
$\chi^\mu = (\frac{1}{4\sqrt{3}}a, \frac{1}{4}a)$ respectively . 
So we found the continuum limit expression for link entropies for
these two directions to be (\ref{45}) and
(\ref{46}) respectively .
If we consider a continuous curve which extends in a possible
direction for an $A$-type curve and parametrize it by 
$y^\mu(s)$, and infinitesimal piece contribute to the sum
(\ref{68}).
The contribution is calculated in the following way:
First expand the infinitesimal $dy^\mu(s)$ on the two directions
\begin{equation}
dy^\mu=(\sqrt{\frac{3}{4}}a, \frac{1}{2}a)dn_{\oO} +
(\sqrt{\frac{3}{4}}a, -\frac{1}{2}a) dn_{\Oo} \label{gthrs63}
\end{equation}
Then the contribution to the sum is
\begin{eqnarray}
d\Sigma'K_{GH}&=& K_{\oO}(x)dn_{\oO} + K_{\Oo}(x)dn_{\Oo}, \\\nonumber
&=&\Bigl\{S(x)-\log Cn R(x) +(\frac{1}{4\sqrt{3}}a, -\frac{1}{4}a)^\rho
\frac{\partial}{\partial x^\rho}\log Cn R(x)\Bigr\}dn_{\oO} \\
&&+\Bigl\{S(x)-\log Cn R(x) 
+(\frac{1}{4\sqrt{3}}a, \frac{1}{4}a)^\rho\frac{\partial}{\partial x^\rho}\log Cn R(x)\Bigr\}dn_{\oO}  
\end{eqnarray}
Clearly
\begin{equation}
dn_{\oO}=\frac{1}{a}(\frac{1}{\sqrt{3}}, 1)_\rho dy^\rho
\end{equation}
and
\begin{equation}
dn_{\Oo}=\frac{1}{a}(\frac{1}{\sqrt{3}}, -1)_\rho dy^\rho
\end{equation}
and so 
\begin{eqnarray}
d\Sigma K_{GH}&=&\bigl\{S-\log CnR(x)+\frac{1}{4\sqrt{3}}a\cdot
\frac{\partial}{\partial x'}\log CnR(x)\bigr\}(dn_{\oO}+dn_{\Oo})\\
&&+\bigl\{-\frac{1}{4}a \frac{\partial}{\partial x^2}\log CnR(x)\bigr\}
\cdot (dn_{\oO}+dn_{\Oo}) \nonumber\\
&=&(S-\log CnR(x)+\frac{1}{4\sqrt{3}}a\frac{\partial}{\partial x^1}\log
CnR(x)\cdot
\frac{1}{a}\frac{2}{\sqrt{3}}dy^1 \nonumber\\
&&+(-\frac{1}{4}a\frac{\partial}{\partial x^2} \log CnR(x)
\frac{2}{a}dy^2\nonumber\\
&=&\frac{\sqrt{3}{4}}A \cdot dn_{\oO} \nonumber\\
dn_{\oO}&=&\frac{4}{\sqrt{3}a}dy^\mu(\frac{1}{2}, \sqrt{\frac{3}{4}})_\mu\\\nonumber
     &=&dy^\mu\frac{1}{a}(\frac{2}{\sqrt{3}},2)_\mu\\\nonumber
&=&\frac{2}{\sqrt{3}a}(S(x)-\log CnR(x))dy^1+\frac{1}{6}dy^1
   \frac{\partial}{\partial x^1}\log CnR(x)\\\nonumber
&& - \frac{1}{2}dy^2\frac{\partial}{\partial x^2}\log CnR(x)
\end{eqnarray}
We know from the fact that the integral of this contribution 
\begin{equation}
\int d\Sigma K_{GH} 
\end{equation}
should integrate up to only depend on the end points - but not on the
way - that the integrability conditions should be satisfied. 
If it were not integrable this way for the curves with an
expansion (\ref{gthrs63}) having $dn_{\oO}$ and $dn_{\Oo}$ both being
positive it would mean that different curves of this type - but with
common end points would not give rise to the same number of classes
of different solution $\varphi$. 
Since the classes of solutions defined from one of these curves are in
a one-to-one correspondence with the ones from the other one there
must be equally many and thus the one-form (\ref{gthrs63}) must be
integrable.

We can rewrite
\begin{eqnarray}
d\sum K_{GH}(S(x)&-&\log Cn R(x)+\frac{1}{4\sqrt3}a
\frac{\partial}{\partial x^1} \log Cn
R(x))\frac{1}{a}\frac{2}{\sqrt3}dy^1 \nonumber\\
+(-\frac{a}{4}\frac{\partial}{\partial
x^2} \log Cn R(x)) \frac{2}{a} dy^2&=&\frac{1}{a}\cdot
\frac{2}{\sqrt3}(S(x)-\log Cn
R(x)+\frac{a}{\sqrt3}\frac{\partial}{\partial x^2}\log Cn
R(x))dy^1 \nonumber\\
&-&\frac{1}{2}d \log Cn R(x).
\end{eqnarray}

In this form the condition for $d\sum K_{GH}$ being a total
differential is very easy to write because the last term
$-\frac{1}{2}d \log Cn R(x)$ itself is one and can be
ignored for that purpose.  Thus we simply get that
\begin{equation}
\frac{\partial}{\partial x^1}(\frac{1}{a}
\cdot\frac{2}{\sqrt3}(S(x)-\log Cn
R(x)+\frac{a}{\sqrt3}\frac{\partial}{\partial x^2} \log Cn R(x))=0
\end{equation}
or that
\begin{equation}\label{79}
f(x^2)=S(x)-\log Cn R(x)+\frac{a}{\sqrt3}\frac{\partial}{\partial
x^2} \log Cn R(x)
\end{equation}
only function of $x^2$ but not of $x^1$.

We recognize this expression (\ref{79}) from the expression occurring in
the form (\ref{49}) or (\ref{59}) for $j^\mu_A(x)$.  We should remember though that in
the derivation leading up to (\ref{49}) or (\ref{59}) we only used the conservation but
did not have the normalization.  Rather we should use our expression
for $d\sum K_{GH}$ which has been physically normalized to
define the correctly normalized $j^\mu_A(x)$ current by
\begin{equation}
d\sum K_{GH}=j^2_{A norm}(x) dx-j^1_{A norm}(x)dx^2
\end{equation}
so that we see
\begin{equation}
j^\mu_{A norm}(x)=\frac{2}{a\sqrt3}j^\mu_A(x).
\end{equation}

One can similarly construct the analogous $d\sum K_{GH}$
differential for the $B$- and the $C$-types of curves and see that the
condition for them being integrable is just equivalent to the current
conservation.  Again we could use this interpretation of the entropy
by means of solution class counting to give the correct normalization
of the current densities,
\begin{eqnarray}
j^\mu_{Anorm}(x)&=&\frac{2}{a\sqrt3}j^\mu_A(x) \nonumber\\
j^\mu_{Bnorm}(x)&=&\frac{2}{a\sqrt3}j^\mu_B(x) \nonumber\\
j^\mu_{Cnorm}(x)&=&\frac{2}{a\sqrt3}j^\mu_C(x) 
\end{eqnarray}

From these considerations it should be understood that our $A$-type
entropy current density $j^2_{Anorm}(x)$, really tells the
logarithm of how many classes of solutions belonging to the
macro scenario are distinguishable by their restriction to one unit
length in the $x^1$-direction.  This is means infinitesimally
that an $dx^1$ is the classification number exp$(j^2_{Anorm
}(x)dx^1)$.  This is indeed the number of states of the fields in the
$dx^1$-interval and thus our entropy current does indeed deserve to be
called an entropy current.

\section{Conclusion and outlook}

We have considered a ``classical'' lattice field theory on a two
dimensional regular triangular lattice of a very general type.  That
is to say we did never write what the supposedly complicated equations
would be.  Rather we just in a extremely abstract manner assumed that we looked
for a big class of solutions obeying all over the lattice some
macro constraints which were meant to be the features to what a
macroscopic observer would notice.  We assumed also in a general way
that the macro scenario imposed corresponding to an adiabatic
development formulated by the assumptions we called 
reversibility in either way.

Our assumption of reversibility or adiabaticity ``either way" was the assumption
(\ref{4.2}) in section 4.
It means that entropy would be constant.
Thus it becomes a setting for studying conserved entropy currents.
It should be understood that such entropy flows are given in a macroscopic
picture which means a system of macro restrictions, consistent with our reversibility.

Now our main study was for such macroscopic pictures or macro scenarios
with our (slightly generalized) adiabaticity assumption to find and
define conserved entropy currents because of the adiabaticity.

The surprise is, however, that instead of finding only one - as
one would presumably have expected - we found \underline{three}
conserved entropy currents, $j^\mu_A(x)$, $j^\mu_B(x)$, and
$j^\mu_C(x)$ as we denoted them in the continuum limit (lattice
constant $a\to 0$).  We can say that we found in our rather general
latticized field theory \underline{three} different kinds of
entropy!

It turned out that with our special simplification of the lattice
model having only restrictions by the field values that are related
inside a certain system of triangles the three different currents of
entropy could be described - in the continuum limit - in terms of two
scalar fields $S(x)$ and $\log Cn R(x)$ describing the prescribed
macro scenario.  From the conservation laws of these three entropies
we got a certain homogeneous third order differential equation
\begin{equation}
\xi^{\mu\nu\rho} \partial_\mu \partial_\nu \partial_\rho \log Cn R(x)=0
\end{equation}
where the third order tensor $\xi^{\mu\nu\rho}$ is related to the
lattice orientation.

What we did by means of the lattice non rotational invariance was to
let the lattice select some direction in which to say that the entropy
flows.  We have these different crude directions $A$, $B$, and $C$
``half-curves''.  In Minkowskian relativistic theories one must use
the distinction between forward in time going and backward going
(half) curves.  
To keep the precise analogy we should though rather
think of right going and left going half-curves.  But we managed
under some more general condition to obtain even several entropy
current definitions.

We do probably best by admitting that we do not fully yet understand
what is going on with these the somewhat mysterious
\underline{three} entropies.  It is rather obvious that as it comes
out these entropies are seemingly lattice artifacts in as far as the
bunches of ``half'' curves extending from some point $x^\mu$ which we
used for the definition of the three entropies were clearly defined by
means of our lattice only.  So from the very definition we had our
three entropies attached to the lattice.

It is our next important subjects of works to study further detailed investigations
of the properties of three entropy currents in two dimensional example and to 
generalize our method of constructing entropy currents into any dimensions.
Furthermore most importantly we will have to clarify the question whether all the
three entropy currents are relevant ones in actual physical situation.

\begin{flushleft}
 \bf\large Acknowledgments
\end{flushleft}

The authors acknowledge the Niels Bohr Institute (Univ. Copenhagen) and 
Yukawa Institute for Theoretical Physics (Kyoto Univ.) for their hospitality 
extended to one of them each.
The work is supported by 
Grand-in-Aids for Scientific Research on Priority Areas, 
Number of Area 763 ``Dynamics of Strings and Fields", 
from the Ministry of Education of Culture, Sports, Science 
and Technology, Japan.

\end{document}